\documentclass[aps,prl,showpacs,twocolumn]{revtex4}
\usepackage{amsmath}
\usepackage{graphicx}

\begin{document}

\title{ Antiferromagnetism in two-dimensional quasicrystals }

\author{ Anuradha Jagannathan and Attila Szallas}
\affiliation{Laboratoire de Physique des Solides, CNRS-UMR 8502, Universit\'e
Paris-Sud, 91405 Orsay, France }

\date{\today}

\begin{abstract}
Quasiperiodic structures possess long range positional order, but are freed of constraints imposed by translational invariance. For spins interacting via Heisenberg couplings, one may expect therefore to find novel magnetic configurations in such structures. We have studied magnetic properties for simple two dimensional models, as a first step towards understanding experimentally studied magnetic quasicrystals such as the Zn-Mg-R(rare earth) compounds. We analyse properties of the antiferromagnetic ground state and magnon excitation modes for bipartite tilings such as the octagonal and Penrose tilings. In the absence of frustration, one has an inhomogeneous Neel-ordered ground state, with local quantum fluctuations of the local staggered magnetic order parameter. We study the spin wave spectrum and wavefunctions of such antiferromagnets within the linear spin wave approximation. Some results for spin-spin correlations in these systems are discussed.
\end{abstract}

\maketitle

Recent experimental investigations of rare-earth based magnetic quasicrystals raise the question of what types of magnetic order are expected to exist in such structures. The Zn-Mg-R series (R: rare-earth) are an example of localized spins in a quasiperiodic geometry, where evidence has been found \cite{sato} for magnetic correlations at low temperatures resulting from predominantly antiferromagnetic interactions between spins. We therefore consider the issue of finding novel quantum magnetic states in this type of system at low temperatures. This question will be addressed here for the simplest theoretical situation, in which there is a spin on every vertex of a perfect quasiperiodic tiling. Study of the experimentally relevant effects of diluting the system and of disorder, which are nontrivial, are reserved for subsequent studies \cite{phasons}. We describe the structure of the ground state and the spectrum of excitations for antiferromagnetically coupled Heisenberg $S=1/2$ spins for this simplest case. At zero temperature, despite the fact that thermal fluctuations are frozen out, the spins do not order perfectly, due to quantum fluctuations. We consider the case of $S=\frac{1}{2}$, the case of greatest theoretical interest, as
quantum effects are expected to be strongest.  

We consider spins located on vertices of two-dimensional tilings: the octagonal tiling formed from squares and $45^\circ$ rhombuses, and the Penrose tiling formed from thick ($72^\circ$) and thin ($36^\circ$) rhombuses.  The eight-fold symmetric octagonal tiling is a somewhat simpler structure compared to the Penrose tiling (five-fold symmetric), in terms of the local environments present. The latter is more closely related to the three
dimensional icosahedral tiling relevant in the description of a large number of
quasicrystalline alloys. In both cases, at T=0, since
there is no frustration in the model, one expects a N\'eel ordered
ground state, with equal and oppositely directed sublattice
magnetizations. 
We have studied the antiferromagnetic ground state and excitations of the quasiperiodic Heisenberg models using linear spin wave (LSW) theory. When possible, the LSW results will be compared with Quantum Monte Carlo (QMC) data that have been obtained for our models. The calculations involve a numerical diagonalization of Hamiltonians for periodic approximants of different sizes, with a finite size scaling analysis for an extrapolation to the infinite quasiperiodic tiling. The approximants of the octagonal tiling were obtained using the method described in \cite{moss}, while the approximants of the Penrose tiling were obtained by a generalization of the method described for the 36 site Taylor approximant in \cite{dunaud}.  

\bigskip
\noindent
{\bf{Model of the quasiperiodic antiferromagnet}}

We consider the nearest-neighbor antiferromagnetic spin-$\frac{1}{2}$ Heisenberg model,
$H=J \sum_{\langle i,j \rangle} {\mathbf S}_i \cdot {\mathbf S}_j$
where $\langle i,j\rangle $ are pairs of sites linked by an edge. The antiferromagnetic coupling $J>0$ is the
the same for all bonds. This model is bipartite, in that sites belong to one of two
sublattices A and B, with a spin on sublattice A coupled
to spins lying on sublattice B and vice-versa. This property ensures that there is no
frustration, i.e., if one considers classical spin variables, the
ground state is one for which all bonds are ``satisfied" -- with all
the A-sublattice spins pointing in one direction and B-sublattice
spins pointing in the opposite direction. 
The site-dependent staggered magnetization is defined by $m_{si} = \langle \epsilon_i S^{z}_i\rangle = S - \delta m_{si}$, where $\epsilon_i$ takes the values $+1$ and $-1$ on the A and B sublattices respectively. The correction term $\delta m_{si}$ arises due to quantum
fluctuations. The global average of $m_{si}$ over all sites $i$ is denoted $m_s$. Assuming that quantum fluctuations are small, one can replace the exact Hamiltonian by a linearized form. This LSW Hamiltonian is then diagonalized using a generalized Bogoliubov transformation (as described in \cite{lswrefs}), and the complete set of eigenvalues and eigenvectors are obtained. The local magnetizations, ground state energy and magnon modes are thus numerically determined.

\begin{figure}[t]
\begin{center}
\includegraphics[scale=0.3]{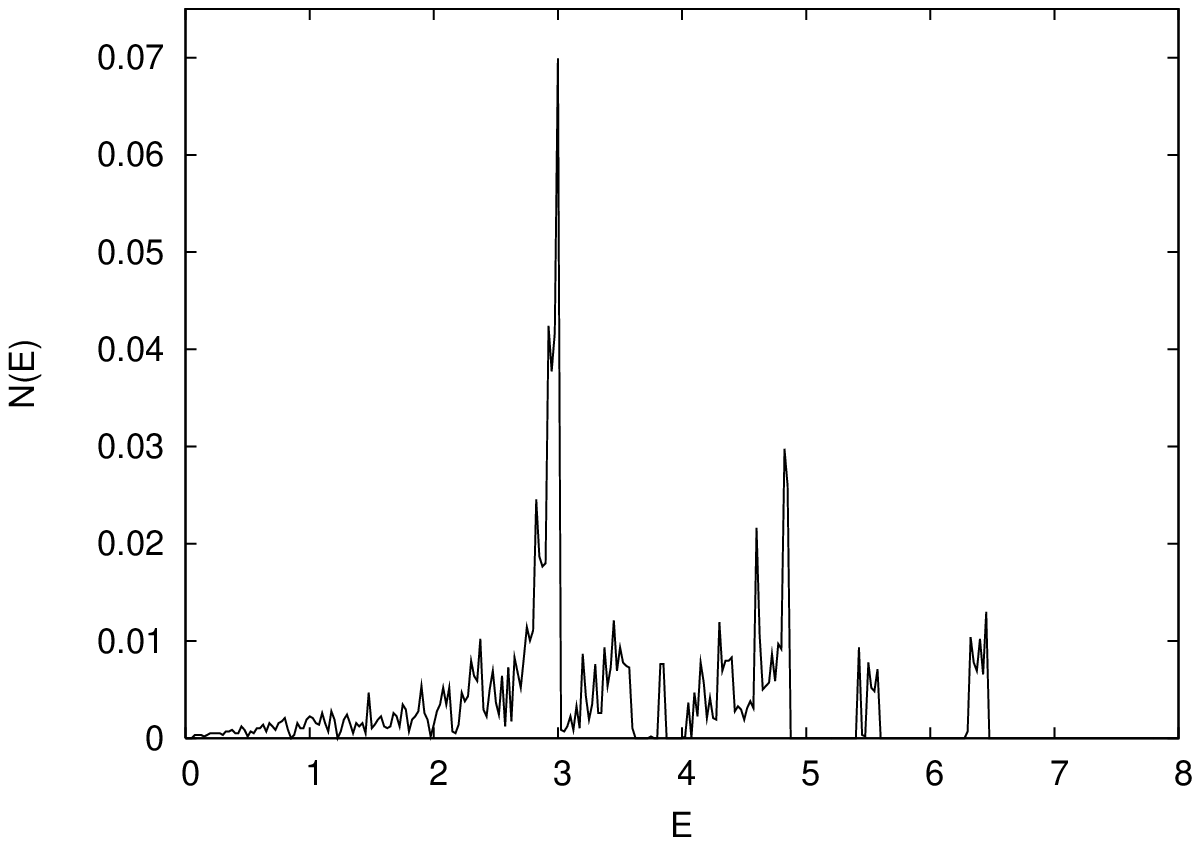}
\hspace{3mm}
\includegraphics[scale=0.3]{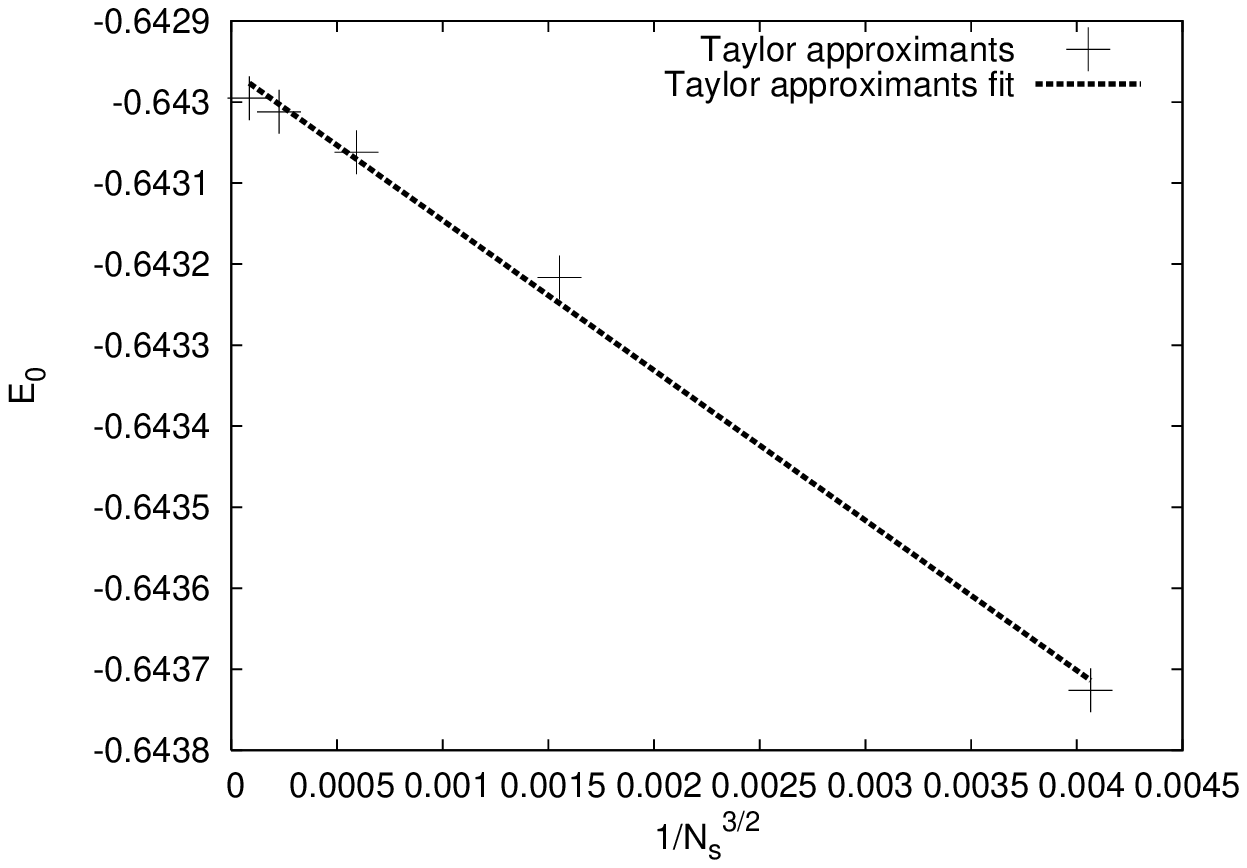}
\caption{(left) Density of states for Taylor $\tau^6$ approximant (N=11556 sites) as a function of energy (expressed in units of J). (right) Scaling of the ground state energy per site $E_0/N$ with system size. }
\end{center}
\end{figure}

\newpage

\bigskip
\noindent
{\bf{The magnon spectrum. Ground state energy}}

The density of states is plotted as a function of the energy (expressed in units of $J$) in Fig.(1)(left) for the case of a Penrose tiling approximant. The main features are: a continuous low energy part, a peak corresponding to degenerate ``loop-like" states at $E=3J$, and high energy bands corresponding to states centered around high coordination number sites. 
The integrated density of states is well-fitted by a quadratic energy dependence, showing that there
is a linear dispersion of the magnon modes in this region of
the spectrum. The effective
spin wave velocity on the Penrose tiling is $c = 1.08 J$, and in the  octagonal tiling, our estimated value is $c_{octa} \approx 1.3 J$.
For comparison , in the square lattice, $c_{sq} =  \sqrt{2}J \approx
1.41 J$ on the square lattice (for edge length $a=1$ and $S=1/2$).
The Penrose tiling spin wave velocity is smaller than in the octagonal 
tiling, which is in turn smaller than the value on the square lattice, reflecting the hierarchy in the areal density of sites. The ground state energy per site is found by extrapolating finite size results as shown in Fig.(1)(right) (the fit to the $N^{-3/2}$ dependence is expected from LSW theory) and has the value
$E_0/N = -0.643(0)$ in units of $J$. This is close to the values found in the octagonal tiling and the square lattice (two cases where analytical calculations are possible), despite spatially different ground states, primarily because the average number of nearest neighbors is the same in all three cases.

\begin{figure}[t]
\begin{center}
\includegraphics[scale=0.3]{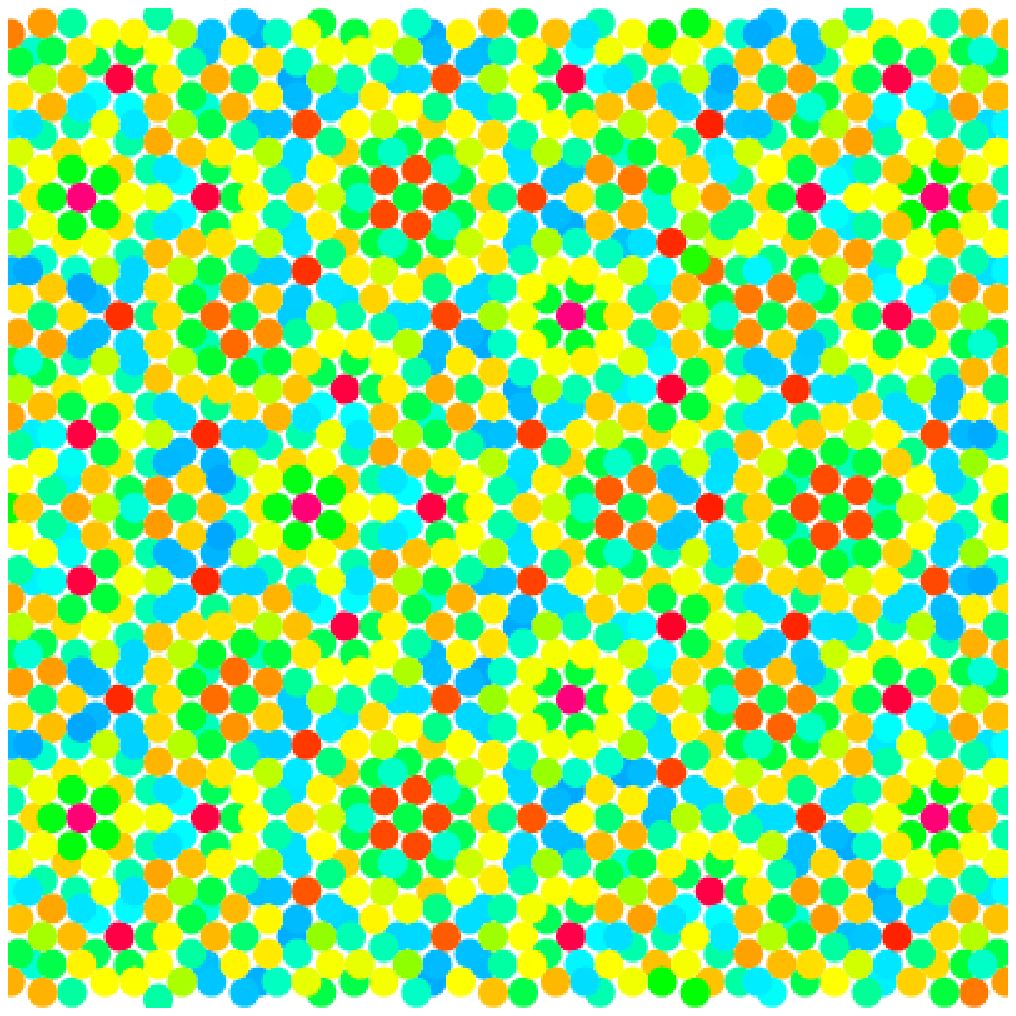}
\hspace{3mm}
\includegraphics[scale=0.3]{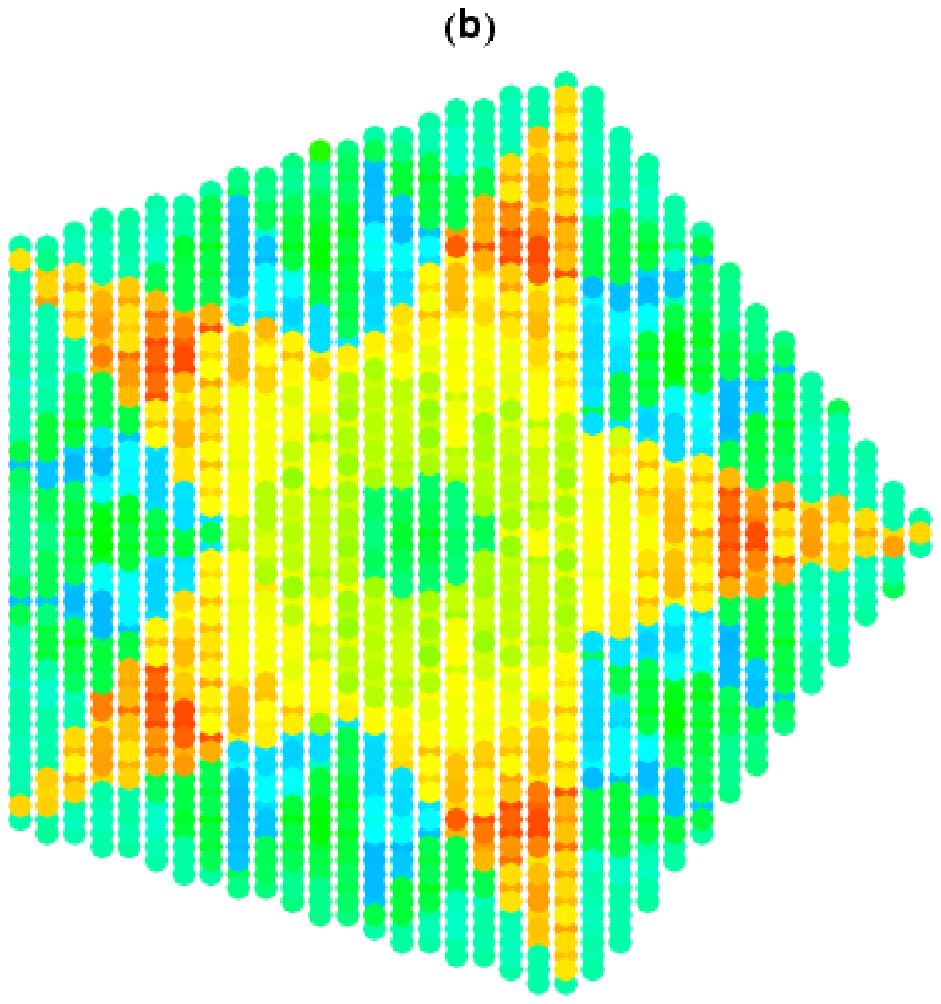}
\caption{(left) Real space distribution of staggered magnetizations on a Penrose approximant (colorscale: smallest(red) to largest(blue)). (right) Staggered magnetizations projected onto the second window in perpendicular space (same color coding) }
\end{center}
\end{figure}

\bigskip
\noindent
{\bf {Wavefunctions. Spatial distribution of the magnetization at T=0} }

Wavefunctions are extended and show multifractal scaling at low energies. One-dimensional states are found at the degenerate energy $E=3J$, while the narrower energy bands at the top of the spectrum correspond to wavefunctions localized on high coordination sites.  From these wavefunctions, one can compute the order parameter in the ground state and study spatial variations. Fig.(2)(left) represents how local magnetizations vary
in space on a portion of the Penrose tiling. The color of the
circles around each vertex varies from red (small magnetization) to
blue (high magnetization). We find that the distribution of magnetizations can be explained qualitatively in terms of a simple local cluster approximation -- the Heisenberg star approximation (see Jagannathan et al in \cite{lswrefs}).  The symmetries hidden in this complex ground state are best shown by representing it in a perpendicular space projection. Fig.(2)(right) shows the magnetizations of subset of sites that project into one of the four pentagon windows in perpendicular space to illustrate this point. One can see the grouping of colors according to the environment. Note: in this representation, the closer two sites are, the farther the distance out to which their environments are identical. 
 
\begin{figure}[t]
\begin{center}
\includegraphics[scale=0.4]{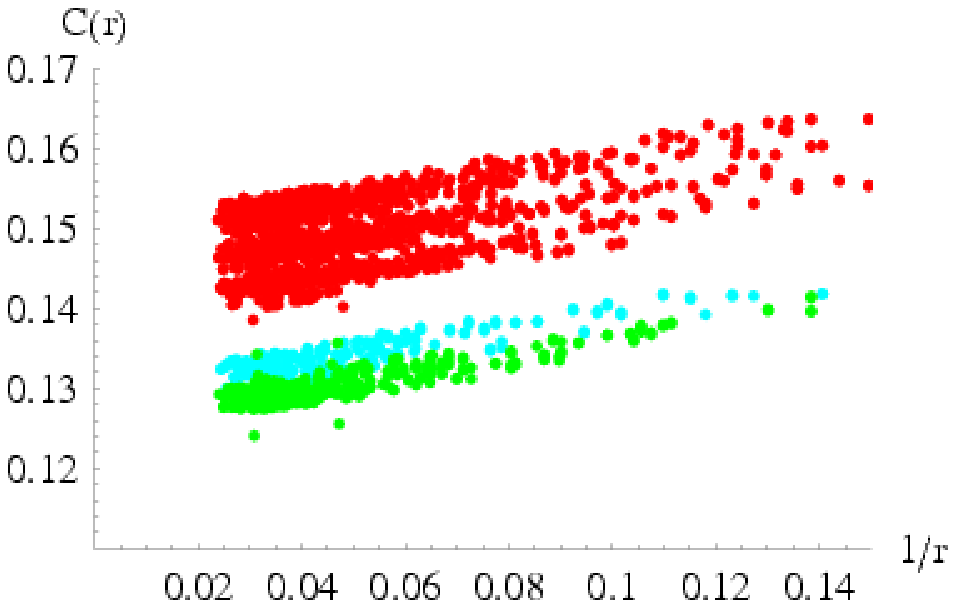}
\hspace{3mm}
\includegraphics[scale=0.3]{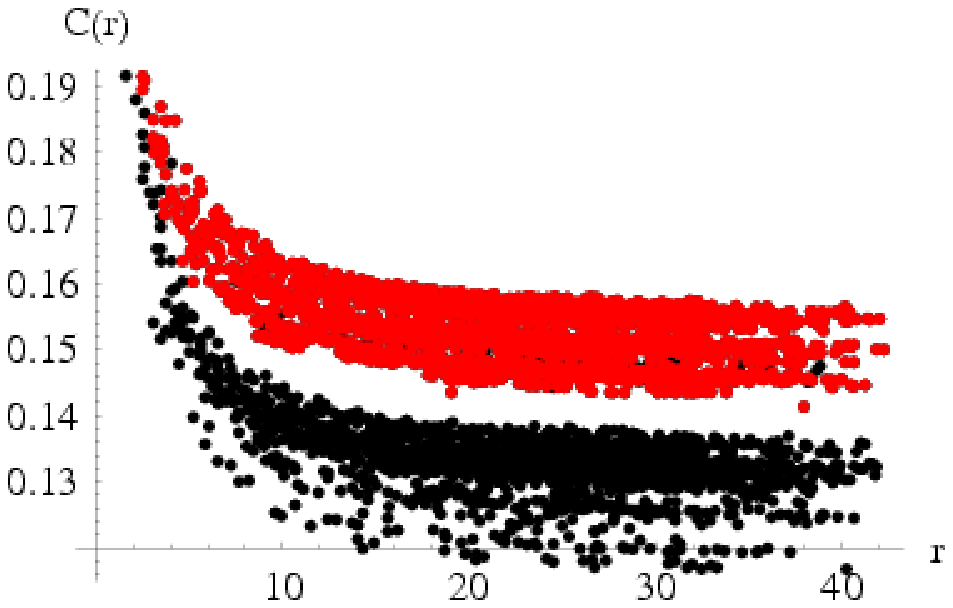}
\caption{(left) Staggered correlation function versus distance from a fixed central $z=3$ site calculated for Taylor $\tau^5$ approximant. (right) The same staggered correlation function versus inverse distance, showing several parallel linear trends (each color represents a different $z$-family) }
\end{center}
\end{figure}
 
\bigskip
\noindent
{\bf Correlation function. Structure factor. } One can define, for a given spin in the tiling $\vec{S}_0$, the staggered spin-spin correlation function $C(\vec{r}) = \langle \epsilon_0 \epsilon_{\vec{r}} \vec{S}_0 \vec{S}_{\vec{r}} \rangle$. As in periodic systems, this function turns out to decay with the inverse of distance $r=\vert\vec{r}\vert$, but it also depends on the type of sites involved. This is shown in the left-hand side figure in Figs.(3), which shows all the spin-spin correlations of a fixed three-fold site placed at the origin in a Taylor $\tau^5$ approximant. The figure shows two groups of sites -- more highly correlated ones (corresponding to the red upper branch) and less correlated ones (the black lower branch). The correlations are grouped in families, each of which decays as the inverse distance from the central site. This is seen more clearly in the right-hand figure, where $C(\vec{r})$ is plotted against $1/r$. Each color corresponds to a site of different coordination number (for example, green points correspond to $z=4$). 
The quasiperiodic character of the antiferromagnetic ground state is also seen in the static structure factor $S(\vec{q})$. Our computations show the apparition of magnetic peaks corresponding to half-integer indices \cite{lswrefs} as expected on general theoretical grounds.

\bigskip
\noindent
{\bf {Conclusions.} } The studies show a rich structure of the magnetic ground state and correlations in 2D quasiperiodic antiferromagnets. Comparison of the average staggered magnetization in the octagonal tiling and in the Penrose tiling shows that the former has larger quantum fluctuations compared to the latter. Spin-spin correlations have a rich structure, with some features of power law decay in common with 2D periodic systems. Interesting questions remain for future investigation, as to quantum coherence and entanglement in such systems, and the effects of perturbations, including phason disorder.

\end{document}